\documentclass[twocolumn,showpacs,10pt]{revtex4-1}%
\usepackage{amsmath,amssymb,graphicx}
\usepackage{amsmath}
\usepackage{amsfonts}
\usepackage{amssymb}
\usepackage[normalem]{ulem}
\usepackage{graphicx}%
\providecommand{\U}[1]{\protect\rule{.1in}{.1in}}
\begin{document}

\title{Two-photon interference imaging}
\author{Deyang Duan}
\email{duandy2015@qfnu.edu.cn}

\affiliation{School of Physics and Physical Engineering, Qufu Normal University, Qufu 273165, China\\
Shandong Provincial Key Laboratory of Laser Polarization and Information
Technology, Research Institute of Laser, Qufu Normal University, Qufu 273165, China}
\author{Yunjie Xia}
\email{yjxia@qfnu.edu.cn}
\affiliation{School of Physics and Physical Engineering, Qufu Normal University, Qufu 273165, China\\
Shandong Provincial Key Laboratory of Laser Polarization and Information
Technology, Research Institute of Laser, Qufu Normal University, Qufu 273165, China}

\begin{quote}
\begin{abstract}
In this article, we propose the two-photon interference imaging based on
two-photon interference mechanism with thermal light source. Theoretical and
experimental results show that the imaging quality and imaging speed of
two-photon interference imaging are comparable to that of classical optical
imaging, and much better than that of conventional quantum imaging (ghost
imaging). Furthermore, Two-photon interference imaging can effectively
overcome the effect of atmospheric turbulence and other harsh optical
environments. The physical essence is the inhibition of two-photon
interference mechanism on atmospheric turbulence.

\end{abstract}
\maketitle
\end{quote}

\section{Introduction}

Since Young completed the double-slit interference experiment in 1807 [1],
interference phenomena have played an important role in fundamental
understanding of photon and have had practical applications [2]. For the
interference of light, Dirac once pointed out that a single photon wave-packet
can only interfere with itself [3], which made it hard to accept for a long
time because the interference was considered as the interference between two
photons. The situation changed in the mid-1980s since the observation of
bi-photon interference of an entangled photon pair generated by spontaneous
parametric down conversion (SPDC) [4]. In the bi-photon interference (or
two-photon interference) [4,5], two photons meet on a beam splitter, their
fourth-order interference can be observed as a coincidence correlation between
two single-photon detectors, each placed at the output ports of the beam
splitter. This two-photon quantum interference effect has been confirmed from
quantum sources, such as SPDC [4], four-wave mixing [6,7], and single photons
from independent sources [8]. Does two-photon interference occur only with
entangled photon pairs? Scarcelli and his coworkers successfully observed the
two-photon interference from two independent chaotic-thermal light sources
[9], which greatly deepens the understanding of two-photon interference. Now,
quantum interference between single photons is one of the most important
physical mechanisms for realizing linear optical quantum computation and
information processing.

In this letter, we propose a new optical imaging technique called two-photon
interference imaging based on the two-photon interference mechanism with
thermal light source. In terms of physical essence, this optical imaging is
still a kind of quantum imaging. Recall, the concept of quantum imaging could
be traced to the pioneering work initiated by Shih \emph{et al} [10], who
realized the first quantum imaging experiment based on the entangled biphoton
generated via SPDC, following the original proposals by Klyshko [11]. Quantum
imaging (or ghost imaging) has not only been demonstrated unique advantages
(e.g., anti-interference [12,13], super-resolution [14]), but also has broad
application prospects (e.g., remote sensing [15], lidar [16,17], pattern
recognition [18]). Not only that, quantum imaging has been beyond the scope of
optical imaging and become a new technology that has important applications in
signal processing [19], atoms [20], and medical fields [21-23]. Although
quantum imaging has lots of aforementioned incomparable advantages and
potential applications [12-23], it seems to be limited in the laboratory
without a breakthrough in commercial applications. Some key issues remain to
be resolved. Efforts to solve these problems have not stopped, but up to now,
the imaging speed and quality of quantum imaging can not be compared with that
of classical imaging.

In conventional quantum imaging setup, the object's image is retrieved by
using two spatially correlated light beams: the reference beam, which never
illuminates the object and is directly measured by a detector with spatial
resolution, and the object beam, which, after illuminating the object is
measured by a bucket detector with no spatial resolution. By correlating the
photocurrents from the two detectors, the image is retrieved. Different from
this, in the scheme of two-photon interference imaging, the light reflected or
transmitted by an object is separated by a beam splitter, the reconstructed
image can be observed by a coincidence correlation between two Charge-coupled
Device (CCD) detectors, each placed at the output ports of the beam splitter.
In this letter, theory and experiments have demonstrated that two-photon
interference imaging has strong abilities of anti-interference and high-speed
imaging. Surprisingly, a clear enough image can be obtained by ten samples,
which is impossible for conventional quantum imaging.

\section{theory}

The setup of two-photon interference imaging is depicted in Fig.1. A laser
beam from a continuous wave laser illuminates a rotating ground glass. Thus, a
large number of random sub-sources that each emitted from a point source are
formed on the surface of the rotating ground glass [24]. The radiation at the
object is the result of a superposition among a large number of these random
sub-fields, $%
{\displaystyle\sum\limits_{m=1}}
E_{m}(\rho_{m},t)$, where $E_{m}(\rho_{m},t)$ represents the field emitted by
the $m$th sub-source. The light reflected by the object is separated into two
beams by a 50:50 beam splitter after propagating in free space. One of the
beams is detected by CCD1, which can be expressed as%

\begin{equation}
E_{1}(\rho_{i1},t)=%
{\displaystyle\sum\limits_{m=1}}
E_{m}(\rho_{m},t)g(\rho_{o})T(\rho_{o})g(\rho_{i1}),
\end{equation}
where, $\rho_{i1}$ and $\rho_{o}$ are the transverse coordinates in the CCD1
plane and the object plane, respectively. $g(\rho_{o})$ is the Green function,
which propagates the $m$th subfield from the $m$th sub-source (coordinate
$\rho_{m}$) to point $\rho_{o}$ on the object plane. $g(\rho_{i1})$ is Green
function, which propagates the flied from $\rho_{o}$ to $\rho_{i}$.
$T(\rho_{o})$ is the function of the object. Correspondingly, the other beam
is detected by CCD2, which can be expressed as
\begin{equation}
E_{2}(\rho_{i2},t)=%
{\displaystyle\sum\limits_{n=1}}
E_{n}(\rho_{n},t)g(\rho_{o})T(\rho_{o})g(\rho_{i2}),
\end{equation}
where $m\neq n$, $\rho_{i2}$ represent the transverse coordinates in the CCD2
plane. In order to simplify the calculation, we have $E_{m,n}^{^{\prime}}%
(\rho_{i1,2},t)=E_{m,n}(\rho,t)g(\rho_{o})T(\rho_{o})g(\rho_{i1,2})$.
\begin{figure}[ptbh]
\centering
\fbox{\includegraphics[width=1\linewidth]{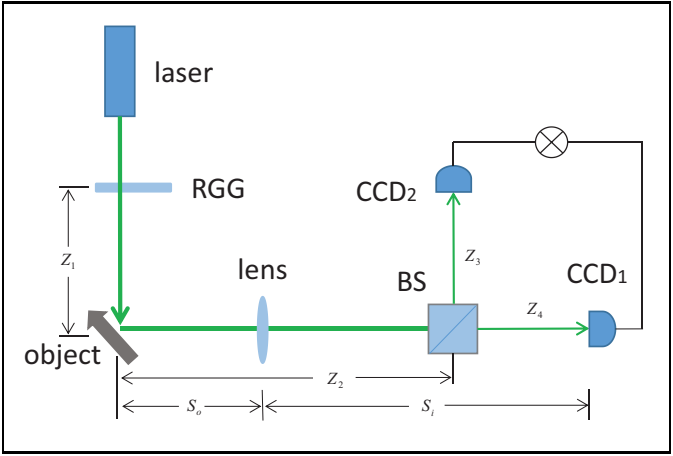}}\caption{(Color online)
The schematic of two-photon interference imaging with thermal light source.
RGG: rotating ground glass, BS: beam splitter.}%
\label{fig:false-color}%
\end{figure}\ \ \

The reconstructed image is observed from the photon number fluctuation
correlation. To calculate the photon number fluctuation correlation, we start
from examining the second-order coherence function $G^{(2)}\left(  \rho
_{i1},\rho_{i2}\right)  $, which is jointly measured by CCD1 and CCD2 on the
two image planes:%
\begin{align}
&  G^{(2)}\left(  \rho_{i1},\rho_{i2}\right) \nonumber\\
&  =\left\langle E_{1}^{\ast}(\rho_{i1})E_{1}(\rho_{i1})E_{2}^{\ast}(\rho
_{i2})E_{2}(\rho_{i2})\right\rangle \nonumber\\
&  =\left\langle \sum_{m}E_{m}^{^{\prime}\ast}(\rho_{i1})\sum_{p}%
E_{p}^{^{\prime}}(\rho_{i1})\sum_{n}E_{n}^{^{\prime}\ast}(\rho_{i2})\sum
_{q}E_{q}^{^{\prime}}(\rho_{i2})\right\rangle \nonumber\\
&  =\sum_{m}\left\vert E_{m}^{^{\prime}}(\rho_{i1})\right\vert ^{2}\sum
_{n}\left\vert E_{n}^{^{\prime}}(\rho_{i2})\right\vert ^{2}\nonumber\\
&  +\sum_{m\neq n}E_{m}^{^{\prime}\ast}(\rho_{i1})E_{n}^{^{\prime}}(\rho
_{i2})E_{n}^{^{\prime}\ast}(\rho_{i2})E_{m}^{^{\prime}}(\rho_{i1})\nonumber\\
&  =\left\langle n(\rho_{i1})\right\rangle \left\langle n(\rho_{i2}%
)\right\rangle +\left\langle \Delta n(\rho_{i1})\Delta n(\rho_{i2}%
)\right\rangle .
\end{align}
The term $\left\langle n(\rho_{i1})\right\rangle \left\langle n(\rho
_{i2})\right\rangle $ corresponds to the product of two identical classical
images measured by CCD1 and CCD2, respectively. The interference term that
generates an image in the joint photon number fluctuation measurement of CCD1
and CCD2.
\begin{align}
&  \left\langle \Delta n(\rho_{i1})\Delta n(\rho_{i2})\right\rangle
\nonumber\\
&  =\sum_{m\neq n}E_{m}^{^{\prime}\ast}(\rho_{i1})E_{n}^{^{\prime}}(\rho
_{i2})E_{n}^{^{\prime}\ast}(\rho_{i2})E_{m}^{^{\prime}}(\rho_{i1})\nonumber\\
&  =\left\vert \sum_{m}E_{m}^{^{\prime}\ast}(\rho_{i1})E_{m}^{^{\prime}}%
(\rho_{i2})\right\vert ^{2}\nonumber\\
&  \equiv\left\vert G_{12}^{(1)}\left(  \rho_{i1},\rho_{i2}\right)
\right\vert ^{2},
\end{align}
where
\begin{align}
&  G_{12}^{(1)}\left(  \rho_{i1},\rho_{i2}\right) \nonumber\\
&  =\sum_{m}E_{m}^{\ast}\int d\rho_{o}g_{m}^{\ast}\left(  \rho_{o}\right)
\int dkO^{\ast}\left(  \rho_{o}\right)  g_{m}^{\ast}\left(  k,\rho_{i1}\right)
\nonumber\\
&  \times\sum_{m}E_{m}\int d\rho_{o^{^{\prime}}}g_{m}\left(  \rho
_{o^{^{\prime}}}\right)  \int dk^{^{\prime}}O\left(  \rho_{o^{^{\prime}}%
}\right)  g_{m}\left(  k^{^{\prime}},\rho_{i2}\right) \nonumber\\
&  =\sum_{m}E_{m}^{\ast}\int d\rho_{o}d\rho_{o^{^{\prime}}}g_{m}^{\ast}\left(
\rho_{o}\right)  E_{m}g_{m}\left(  \rho_{o^{^{\prime}}}\right) \nonumber\\
&  \times\int dkO^{\ast}\left(  \rho_{o}\right)  e^{ik\rho_{o}}somb\left(
\frac{\pi}{\lambda}\frac{D}{s_{o}}\left\vert \rho_{o}+\frac{\rho_{i1}}{\mu
}\right\vert \right) \nonumber\\
&  \times\int dk^{^{\prime}}O\left(  \rho_{o^{^{\prime}}}\right)
e^{-ik^{^{\prime}}\rho_{o^{^{\prime}}}}somb\left(  \frac{\pi}{\lambda}\frac
{D}{s_{o}}\left\vert \rho_{o^{^{\prime}}}+\frac{\rho_{i2}}{\mu}\right\vert
\right)
\end{align}
is the first-order coherence function. $k$ and $\lambda$ represent wave vector
and wavelength respectively. $D$ is the diameter of the imaging lens.
$\mu=s_{i}/s_{o}$ is the magnification factor. $s_{o}$ is the distance between
the object and the imaging lens, $s_{i}$ is the distance between the imaging
lens and the image plane. $\left\langle T(\rho_{o})T^{\ast}(\rho
_{o})\right\rangle =\lambda O(\rho_{o})\delta\left(  \rho_{o}-\rho
_{o}^{^{\prime}}\right)  $. For a perfect imaging system, the $somb$ function
(or point-spread function) in the convolution of Eq.5 will be replaced by $\delta$
function. However, limited by the finite size of the imaging system, we may
never have a perfect point-to-point relationship.

To calculate Eq.5, we complete the summation in terms of the sub-sources by
means of an integral over the entire source plane. Thus, the Eq.5 is
approximated in the follow form,%

\begin{align}
&  G_{12}^{(1)}\left(  \rho_{i1},\rho_{i2}\right)  \nonumber\\
&  \propto\int d\rho_{o}\left\vert O(\rho_{o})\right\vert ^{2}somb\left(
\frac{\pi}{\lambda}\frac{D}{s_{o}}\left\vert \rho_{o}+\frac{\rho_{i1}}{\mu
}\right\vert \right)  \nonumber\\
&  \times somb\left(  \frac{\pi}{\lambda}\frac{D}{s_{o}}\left\vert \rho
_{o}+\frac{\rho_{i2}}{\mu}\right\vert \right)  .
\end{align}
Submitting Eq.6 into Eq.4, we have%

\begin{align}
&  \left\langle \Delta n(\rho_{i1})\Delta n(\rho_{i2})\right\rangle
\nonumber\\
&  \propto\left\vert \int d\rho_{o}\left\vert O(\rho_{o})\right\vert
^{2}somb^{2}\left(  \frac{\pi}{\lambda}\frac{D}{s_{o}}\left\vert \rho
_{o}+\frac{\rho_{i1}}{\mu}\right\vert \right)  \right. \nonumber\\
&  \times\left.  somb^{2}\left(  \frac{\pi}{\lambda}\frac{D}{s_{o}}\left\vert
\rho_{o}+\frac{\rho_{i2}}{\mu}\right\vert \right)  \right\vert ^{2}.
\end{align}
Equation (7) shows that the reconstructed image can be observed by measuring
the term of two-photon interference. Moreover, we will obtain a perfect image
when $\rho_{i1}=\rho_{i2}$.

A significant advantage of two-photon interference imaging is that it can
effectively overcome the effect of harsh optical environment, e.g. atmospheric
turbulence. Next, we illustrate this feature. In order to simplify the
calculation, we assume that the atmospheric turbulence is introduced between
the object and the beam splitter. Thus, the light fields received by two CCD
cameras can be expressed as%
\begin{align}
E_{1}(\rho_{i1},t)  & =\sum_{m}E_{m}(\rho_{m},t)g(\rho_{o})T(\rho_{o}%
)g(\rho_{i1})e^{i\delta\varphi_{m}}\nonumber\\
& =E_{m}^{^{\prime\prime}}(\rho_{i1},t)e^{i\delta\varphi_{m}};
\end{align}

\begin{align}
E_{2}(\rho_{i2},t) &  =\sum_{n}E_{n}(\rho_{n},t)g(\rho_{o})T(\rho_{o}%
)g(\rho_{i2})e^{i\delta\varphi_{n}}\nonumber\\
&  =E_{n}^{^{\prime\prime}}(\rho_{i2},t)e^{i\delta\varphi_{n}}.
\end{align}
The function $e^{i\delta\varphi}$ stands for the atmospheric turbulence
induced phase variations. According to the above calculation process, the
photon number fluctuation correlation can be expressed as%

\begin{align}
&  G^{(2)}\left(  \rho_{i1},\rho_{i2}\right) \nonumber\\
&  =\left\langle \sum_{m}E_{m}^{^{^{\prime\prime}}\ast}(\rho_{i1}%
)e^{i\delta\varphi_{m}}\sum_{p}E_{p}^{^{^{\prime\prime}}}(\rho_{i1}%
)e^{i\delta\varphi_{p}}\right. \nonumber\\
&  \left.  \sum_{n}E_{n}^{^{^{\prime\prime}}\ast}(\rho_{i2})e^{i\delta
\varphi_{n}}\sum_{q}E_{q}^{^{^{\prime\prime}}}(\rho_{i2})e^{i\delta\varphi
_{q}}\right\rangle \tag{10(a)}\\
&  =\sum_{m}\left\vert E_{m}^{^{^{\prime\prime}}}(\rho_{i1})e^{i\delta
\varphi_{m}}\right\vert ^{2}\sum_{n}\left\vert E_{n}^{^{^{\prime\prime}}}%
(\rho_{i2})e^{i\delta\varphi_{n}}\right\vert ^{2}\nonumber\\
&  +\sum_{m\neq n}E_{m}^{^{^{\prime\prime}}\ast}(\rho_{i1})E_{n}%
^{^{^{\prime\prime}}}(\rho_{i2})E_{n}^{^{^{\prime\prime}}\ast}(\rho_{i2}%
)E_{m}^{^{^{\prime\prime}}}(\rho_{i1})\nonumber\\
&  \times e^{i\delta\varphi_{m1}}e^{i\delta\varphi_{n1}}e^{i\delta\varphi
_{m2}}e^{i\delta\varphi_{n2}}\tag{10(b)}\\
&  =\left\langle n^{^{\prime}}(\rho_{i1})\right\rangle \left\langle
n^{^{\prime}}(\rho_{i2})\right\rangle +\left\langle \Delta n^{^{\prime}}%
(\rho_{i1})\Delta n^{^{\prime}}(\rho_{i2})\right\rangle . \tag{10(c)}%
\end{align}
The first two terms of Eq.10(c) represent the classical images output by two
CCD cameras, which cannot eliminate the influence of turbulence. However, the
cross interference term $\left\langle \Delta n(\rho_{i1})\Delta n(\rho
_{i2})\right\rangle $ reaches its turbulence-free when $z_{3}=z_{4}$,
$e^{i\delta\varphi_{m1}}e^{i\delta\varphi_{n1}}e^{i\delta\varphi_{m2}%
}e^{i\delta\varphi_{n2}}=1$.

\section{Experiments}

\begin{figure}[ptbh]
\centering
\fbox{\includegraphics[width=1\linewidth]{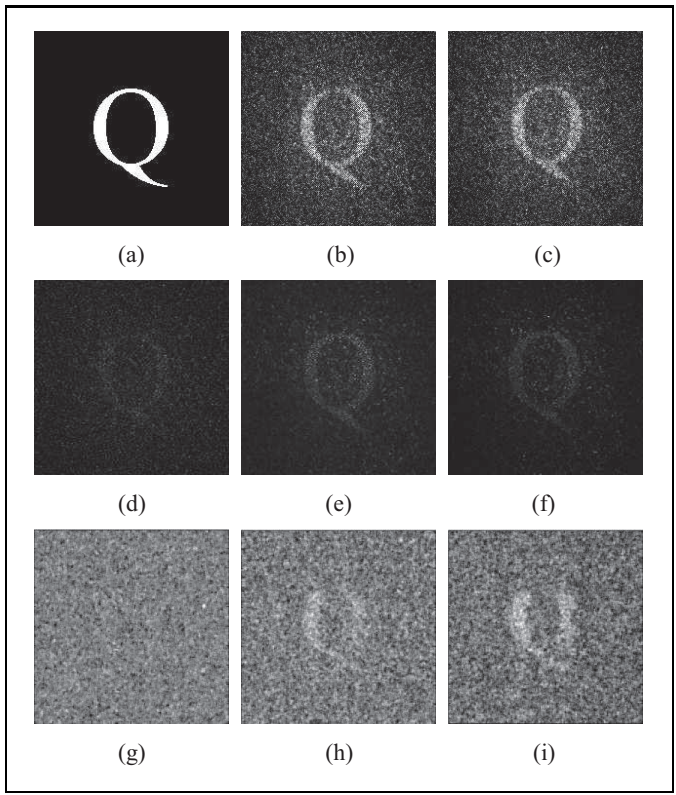}}\caption{Top row: (a)
object (letter ``Q''), (b) classic image output by CCD1, (c) classic image
output by CCD2 (image has been flipped). Middle row: two-photon interference
image obtained by $\left\langle \Delta n(\rho_{i1})\Delta n(\rho
_{i2})\right\rangle $ with different frames. The numbers of frames are (d) 10,
(e) 500, (f) 1000. Bottom row: the corresponding quantum image with different
frames. The numbers of frames are (g) 10, (h) 500, (i) 1000.}%
\label{fig:false-color}%
\end{figure}

The experimental setup is schematically shown in Fig.1. In the setup, a
standard monochromatic laser (30 mW) with wavelength $\lambda=532$nm
illuminates a rotating ground glass (2$rad/$min). Thus, millions of tiny
diffusers within the rotating ground glass scatter the laser beam into many
independent wave packets, which generate a typical chaotic pseudothermal
source with fairly large size in transverse dimension. After free propagation
of $z_{1}=100$cm, the light beam illuminates the object, such as the letter
\textquotedblleft Q\textquotedblright\ as shown in Fig.2(a). The scattered and
reflected photons reflected by the object is separated into two beams by a
50:50 beam splitter after propagation of $z_{2}\simeq103$cm. one of the beam
is collected by CCD1 (the imaging source, DFK 23U618) and the other is
collected by CCD2 (the imaging source, DFK 23U618). An imaging lens focuses
the scattered and reflected light from the object onto the two image planes
(CCD plane) defined by the Gaussian thin lens equation $1/s_{o}+1/s_{i}=1/f$,
where $s_{o}=60$ and $s_{i}\simeq43$ represent object distance and image
distance, respectively. $f=25$cm is the focal length of the imaging lens.

Two CCD cameras are controlled by software to collect data at the same time. A
photon number fluctuation correlation circuit [25-27] is used to measure the
photon number fluctuation correlation. Figure 2 reports a set of typical
experimental results. Figure 2(b) and Figure 2(c) are two classic images output by the two CCD
cameras in two-photon interference experiment setup, i.e., $\left\langle n(\rho
_{i1})\right\rangle \left\langle n(\rho_{i2})\right\rangle $. Figure 2(d-f) is
a measurement of the cross interference term $\left\langle \Delta n\left(
\rho_{i1}\right)  \Delta n\left(  \rho_{i2}\right)  \right\rangle $. It is
this cross interference term that generates an image in the joint photon
number fluctuation measurements of CCD1and CCD2. Figure 2(g-i) is the
experimental results of conventional quantum imaging (ghost imaging). In the
setup of quantum imaging, all the devices and parameters are the same as the
two-photon interference imaging experiment. From Fig.2 we obtain the following
conclusion: (i) The imaging quality of two-photon interference imaging is much
higher than that of quantum imaging, especially when the number of samples is
very small. Two photon interference imaging can produce a clear image by ten
samples, which is impossible for conventional quantum imaging. The reason is
that two-photon interference imaging measures the spatial resolution of both
optical paths, while conventional quantum imaging only measures the spatial
resolution of one optical path, the other optical path is measured by point
measurement. The imaging quality of two-photon interference imaging is
slightly lower than classical imaging. (ii) The imaging speed of two-photon
interference imaging is much faster than that of quantum imaging and can be
compared with that of classical optical imaging. \begin{figure}[ptbh]
\centering
\fbox{\includegraphics[width=1\linewidth]{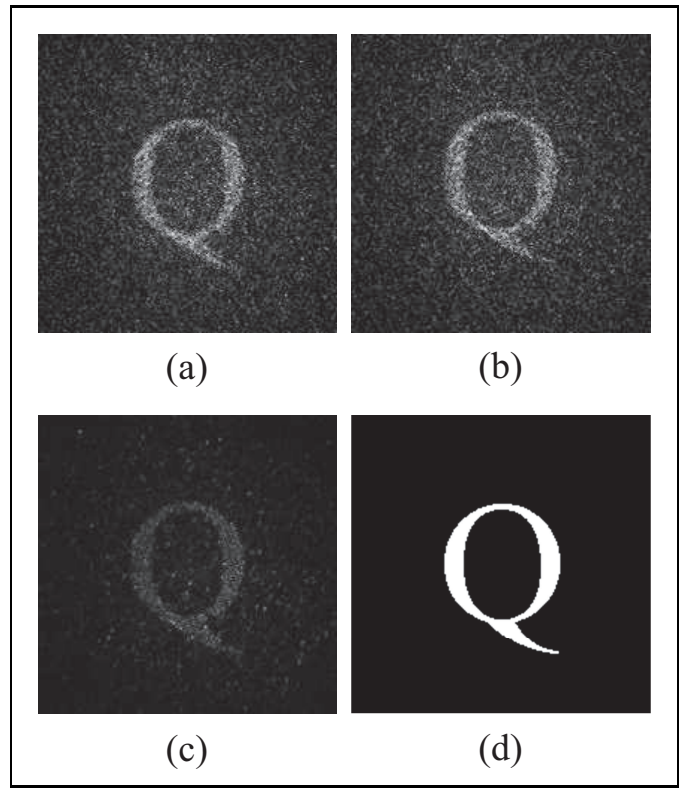}}\caption{(a,b)Two classical
images of the letter \textquotedblleft Q\textquotedblright\ showing the
distortion caused by turbulence. The shape and position of the image are changed
by turbulence, which results in image distortion. (c)Experimentally generated turbulence-free two-photon interference
images of a letter \textquotedblleft Q\textquotedblright\ object using 200
frames of data measured through laboratory turbulence. (d) Object.}%
\label{fig:false-color}%
\end{figure}

Next, we demonstrate the turbulence-free feature of two-photon interference
imaging. In this experiment, atmospheric turbulence is introduced to the
optical paths between the object and lens by adding heating elements
underneath the optical paths operating at a temperature of 550%
${}^{\circ}{\rm C}$
with the refractive index structure parameter $C_{n}^{2}$ in the range of
$1.5\times10^{-10}$ to $1.0\times10^{-9}$ [13]. The length of the heating area
is $20$ cm. These values correspond to extremely high levels of atmospheric
turbulence causing significant temporal and spatial fluctuations of light
intensity. In the measurement, the two-photon interference image and classical
image of the object were captured and monitored simultaneously when the
turbulence was introduced to optical paths. The observations are reported in
Fig.3. The experimental results show that two-photon interference imaging can
effectively overcome the influence of atmospheric turbulence.

\section{Conclusion}

In summary, the two-photon interference imaging based on thermal light source
has been demonstrated in this article. Theoretical and experimental results
show that the imaging quality and imaging speed of the two-photon interference
imaging is much better than that of conventional quantum imaging, and
comparable to classic optical imaging. The reason is that both optical paths
carry the information of the object and are measured with spatial resolution.
More important, Two photon interference imaging can effectively overcome the
effect of harsh optical environment. Further more, two photon interference
imaging has other advantages, for example, its optical structure is similar to
classical optical camera, which makes it suitable for practical application.

\section{Acknowledgement}

This project was supported by the National Natural Science Foundation (China)
under Grant Nos. 11704221, 11574178 and 61675115, Taishan Scholar Project of
Shandong Province (China) under Grant No. tsqn201812059.

\end{document}